\documentclass[aps,prb,twocolumn,superscriptaddress,longbibliography,floatfix]{revtex4-2} 
\usepackage{amsmath,amssymb,graphicx,bm,mathtools}
\usepackage[colorlinks=true,citecolor=blue,linkcolor=blue,urlcolor=blue]{hyperref}
\usepackage{subcaption}
\usepackage{tikz}
\usepackage{float}
\usepackage{pgfplots}
\pgfplotsset{compat=1.18}

\newcommand{\PT}{\mathcal{PT}}

\begin{document}

\title{Moir\'e-Enhanced Plasmonics in Non-Hermitian Twisted Bilayer Graphene}

\author{Andrianos Sygrimis}
\email[Corresponding author: ]{asygrimis@physics.uoc.gr}
\affiliation{Department of Physics, University of Crete, 70013 Heraklion, Greece}
\affiliation{Institute of Electronic Structure and Laser (IESL), FORTH, 71110 Heraklion, Greece}
\author{Giorgos P. Tsironis}
\affiliation{Department of Physics, University of Crete, 70013 Heraklion, Greece}
\affiliation{Institute of Electronic Structure and Laser (IESL), FORTH, 71110 Heraklion, Greece}
\affiliation{Department of Physics, Harvard University, Cambridge, Massachusetts 02138, USA}

\begin{abstract}
We study plasmonic excitations in twisted bilayer graphene within a non-Hermitian framework that incorporates effective gain and loss. Using a non-Hermitian extension of the Bistritzer--MacDonald continuum model together with a biorthogonal Kubo formalism for the optical conductivity, we determine how the moir\'e electronic structure enters the plasmonic response of the active bilayer. We find that non-Hermiticity modifies the collective spectrum, yielding optical and acoustic plasmon branches, with the acoustic branch exhibiting strong subwavelength confinement. In the parity-time-symmetric configuration, gain--loss engineering can reduce the effective spatial damping and enhance the propagation length within the ideal linear model. The same regime produces strongly localized transverse-magnetic near fields. We argue that the enhancement is not a generic consequence of adding gain to a bilayer, but results from the combined influence of moir\'e-band reconstruction, biorthogonal optical matrix elements, and non-Hermitian modification of the plasmon pole. We also discuss the limitations imposed by disorder, substrate loss, gain saturation, and stability of the parity-time-symmetric regime. These results identify twisted bilayer graphene as a promising, but experimentally demanding, platform for tunable non-Hermitian plasmonics in moir\'e quantum materials.
\end{abstract}

\maketitle

\section{Introduction}

Twisted bilayer graphene (TBG) has emerged as one of the central platforms in moir\'e quantum matter, owing to the strong reconstruction of its low-energy electronic structure under a relative twist between two graphene sheets. Near the magic angle, the hybridization of Dirac cones produces ultra-flat moir\'e bands, drastically reducing the electronic kinetic energy and amplifying the role of interactions \cite{bistritzer2011moire,tarnopolsky2019origin,carr2019exact,andrei2020graphene}. This mechanism underlies the discovery of correlated insulating states and unconventional superconductivity in magic-angle graphene superlattices, and has established TBG as a versatile laboratory for studying topology, symmetry breaking, and strongly correlated phases in two-dimensional materials \cite{cao2018correlated,cao2018unconventional,isobe2018unconventional,po2018origin,koshino2018maximally,shavit2021theory,oh2021evidence}. Beyond its electronic many-body phenomena, the moir\'e tunability of TBG also offers a promising route toward engineering collective excitations and optical response in ways that are inaccessible in conventional crystalline systems.

In parallel, graphene plasmonics has developed into a mature framework for manipulating electromagnetic energy at deeply subwavelength scales. Owing to the massless Dirac spectrum, gate-tunable carrier density, and strong field confinement, graphene supports plasmonic modes spanning the terahertz to mid-infrared regime with remarkable electrostatic tunability \cite{jablan2009plasmonics,koppens2011graphene,grigorenko2012graphene,low2014graphene,goncalves2016grapheneplasm,maier2007plasmonics}. Experimental advances such as infrared nano-imaging and the realization of highly confined low-loss plasmons in graphene-based heterostructures have further established two-dimensional materials as a natural arena for flatland plasmonics \cite{fei2012gate,chen2012optical,woessner2015highly,basov2016polaritons,chen2017flatland}. Because plasmonic excitations are exceptionally sensitive to both band structure and dissipation, moir\'e materials such as TBG are expected to host rich collective electromagnetic behavior, where twist angle, filling, screening environment, and interlayer coupling can all reshape the plasmonic landscape.

The use of TBG is therefore not only a geometrical variant of ordinary bilayer graphene. In an untwisted or otherwise generic bilayer, optical and acoustic plasmons may already arise from the symmetric and antisymmetric charge oscillations of the two conducting sheets. In TBG, however, the twist produces a moir\'e mini-Brillouin zone, reconstructed low-energy bands, enhanced density of states near flat or weakly dispersive bands, and optical matrix elements that differ from those of a conventional bilayer. These features enter the conductivity kernel and can modify both the plasmon dispersion and its damping. The purpose of the present work is to isolate this moir\'e-sensitive plasmonic response in the presence of effective gain and loss, rather than to attribute all enhancement to gain compensation alone.

A particularly compelling direction is to examine these collective modes through the lens of non-Hermitian physics. Non-Hermitian Hamiltonians provide an effective description for open systems with gain, loss, leakage, radiative damping, and finite lifetime effects, and they can exhibit phenomena with no Hermitian counterpart, such as parity-time ($\PT$) symmetry breaking, complex spectral bifurcation, and exceptional points \cite{bender1998real,bender2002complex,elganainy2018nonhermitian,ashida2020nonhermitian,bergholtz2021exceptional}. In $\PT$-symmetric settings, the competition between balanced gain and loss gives rise to a transition between an exact phase with real eigenvalues and a broken phase where eigenvalues become complex conjugate pairs \cite{elganainy2007theory,makris2008beam,guo2009observation,ruter2010observation}. More generally, exceptional points correspond to non-Hermitian spectral degeneracies where both eigenvalues and eigenvectors coalesce, generating singular topology, nontrivial mode exchange, and unusual dynamical response \cite{berry2004physics,dembowski2001experimental,heiss2012physics,miri2019exceptional,ozdemir2019parity}. These ideas have reshaped modern photonics and wave physics, with consequences for lasing, energy transfer, sensing, and mode selection \cite{feng2014single,hodaei2014parity,peng2014parity,doppler2016dynamically,xu2016topological,wiersig2014enhancing,hodaei2017enhanced,chen2017exceptional}.

Non-Hermitian effects are especially relevant to plasmonic systems, where intrinsic ohmic damping, radiative decay, and modal hybridization naturally render the effective dynamics non-conservative. Recent studies have shown that plasmonic and graphene-based photonic structures can host exceptional points, $\PT$-symmetry breaking, and loss-enabled functionalities, demonstrating that dissipation is not merely detrimental but can serve as a design resource \cite{alaeian2014nonhermitian,lin2016loss,zhang2017tailoring,chen2019tailoring,li2019chiral,min2020epplasmonic,park2021accessing,zhu2021designing,tayebi2021nonhermitian,ma2021metasurfaces,butler2023ptplasmonic}. Recent near-field experiments and proposals have also connected exceptional-point physics with directional routing of hyperbolic polaritons, illustrating that non-Hermitian degeneracies can have direct consequences for nanoscale polariton propagation \cite{guo2024exceptionalrouting}. These developments suggest that two-dimensional plasmonic platforms can support non-Hermitian spectral engineering with strong confinement, electrical tunability, and potentially enhanced sensitivity to microscopic band-structure features. In this sense, graphene and related van der Waals systems provide an attractive bridge between non-Hermitian photonics and quantum materials.

Very recently, non-Hermitian band topology has also begun to be explored directly in TBG itself. Theoretical works have identified exceptional topology in non-Hermitian twisted bilayer graphene, non-Hermitian band topology in TBG aligned with hexagonal boron nitride, and the concept of exceptional magic angles, indicating that the moir\'e flat-band platform can support fundamentally new non-Hermitian spectral structures \cite{huang2025exceptional,bera2025nhband,esparza2025exceptional}. These results motivate a broader question: how are collective charge oscillations and plasmonic modes modified when the underlying moir\'e electronic environment is treated beyond the Hermitian limit? Since plasmons probe the poles of the density response and are directly shaped by both band dispersion and damping, they are a natural observable through which to investigate exceptional points, $\PT$-related phase transitions, and broken-phase behavior in moir\'e materials.

In this work, we study non-Hermitian plasmonics in twisted bilayer graphene. Our central objective is to understand how the moir\'e electronic structure of TBG, together with non-Hermitian effects associated with dissipation, mode coupling, and spectral broadening, reshapes plasmonic resonances and their attenuation. More specifically, we establish a framework in which the plasmonic response of TBG can be analyzed using non-Hermitian concepts, thereby connecting graphene plasmonics, moir\'e flat-band physics, and $\PT$/exceptional-point phenomena. We do not assume that every enhancement is caused by an exact exceptional point; rather, we examine how proximity to $\PT$-symmetry breaking and the resulting complex spectral reconstruction can modify the conductivity and the plasmon pole. By doing so, we seek to clarify whether TBG can serve not only as a strongly correlated electronic material, but also as a tunable non-Hermitian plasmonic platform with distinctive collective behavior and potential functionality for nanoscale photonics and quantum materials research.
\section{Non-Hermitian Model}

\subsection{Non-Hermitian continuum Hamiltonian}

We describe twisted bilayer graphene using a non-Hermitian extension of the Bistritzer--MacDonald continuum model. For valley $\xi=\pm1$, the effective Hamiltonian is written as

\begin{equation}
H_{\xi,\mathrm{NH}}(\mathbf{k},\mathbf{r}) =
\begin{pmatrix}
h_{\xi,+\theta/2}(\mathbf{k}) + i\lambda_t\sigma_z & T_\xi(\mathbf{r}) \\
T_\xi^\dagger(\mathbf{r}) & h_{\xi,-\theta/2}(\mathbf{k}) + i\lambda_b\sigma_z
\end{pmatrix},
\label{eq:HNH}
\end{equation}
where $h_{\xi,\pm\theta/2}$ are the Dirac Hamiltonians of the two graphene layers rotated by $\pm\theta/2$, while $T_\xi(\mathbf{r})$ denotes the moir\'e-periodic interlayer tunneling matrix. We use
\begin{equation}
h_{\xi,\phi}(\mathbf{k})
=
-\hbar v_F
\left[
R_{\phi}\left(\mathbf{k}-\mathbf{K}_{\xi}^{\phi}\right)
\right]\cdot
\left(\xi\sigma_x,\sigma_y\right),
\label{eq:dirac}
\end{equation}
where $R_{\phi}$ rotates momenta by angle $\phi$, $v_F$ is the monolayer graphene Fermi velocity, and $\sigma_i$ act in the sublattice basis. The moir\'e tunneling is
\begin{equation}
T_\xi(\mathbf{r})
=
\sum_{j=1}^{3} T_{j,\xi}\,
\exp(i\xi\mathbf{q}_j\cdot\mathbf{r}),
\label{eq:tunnel}
\end{equation}
with
\begin{equation}
T_{j,\xi}
=
\begin{pmatrix}
w_0 & w_1 e^{-i\xi\varphi_j}\\
w_1 e^{i\xi\varphi_j} & w_0
\end{pmatrix},
\qquad
\varphi_j=\frac{2\pi(j-1)}{3}.
\label{eq:tunnelmatrix}
\end{equation}
Here $w_0$ and $w_1$ denote AA and AB/BA interlayer tunneling amplitudes, respectively. This explicit form makes clear which ingredients are specific to the twisted structure: the moir\'e wave vectors $\mathbf{q}_j$, the mini-Brillouin zone, and the reconstructed low-energy spectrum generated by interlayer hybridization.

The terms $i\lambda_t\sigma_z$ and $i\lambda_b\sigma_z$ are treated as effective non-Hermitian self-energies. Because they multiply $\sigma_z$, they represent a sublattice-staggered imaginary potential within each layer rather than a microscopic model of a particular pump or reservoir. Positive and negative imaginary parts correspond, in the linearized description, to gain and loss channels. In the ideal symmetric model, the choice $\lambda_t=-\lambda_b$ gives a balanced $\PT$-symmetric gain--loss profile, whereas $\lambda_t\neq-\lambda_b$ describes an imbalanced non-Hermitian environment. This phenomenological description is intended to capture how controlled gain/loss modifies the optical response; a microscopic treatment of a specific active medium would require additional reservoir dynamics beyond the present model.

Because $H_{\xi,\mathrm{NH}}\neq H_{\xi,\mathrm{NH}}^\dagger$, its eigenvalues are generally complex and its right and left eigenvectors are distinct. They satisfy
\begin{align}
H_{\xi,\mathrm{NH}} |R_n\rangle &= E_n |R_n\rangle,\\
H_{\xi,\mathrm{NH}}^\dagger |L_n\rangle &= E_n^* |L_n\rangle,\\
\langle L_m|R_n\rangle &= \delta_{mn}.
\end{align}
All response functions below are therefore evaluated in this biorthogonal basis.

\subsection{Biorthogonal Kubo conductivity}

The optical conductivity is obtained from the current-current response function generalized to a non-Hermitian basis. The total conductivity is separated into intra- and interband parts,
\begin{equation}
\sigma_{\alpha\beta}(\omega)
=
\sigma_{\alpha\beta}^{\mathrm{intra}}(\omega)
+
\sigma_{\alpha\beta}^{\mathrm{inter}}(\omega),
\end{equation}
where $\alpha,\beta\in\{x,y\}$. The velocity matrix elements are
\begin{equation}
V_{mn}^{(\alpha)}
=
\langle L_m|\hat v_\alpha|R_n\rangle,
\qquad
\hat v_\alpha=
\frac{1}{\hbar}
\frac{\partial H_{\xi,\mathrm{NH}}}{\partial k_\alpha}.
\end{equation}

The interband contribution is
\begin{widetext}
\begin{align}
\sigma_{\alpha\beta}^{\mathrm{inter}}(\omega)
&=
\frac{e^2\hbar}{i(2\pi)^2}
\int_{\mathrm{BZ}}d^2k
\sum_{m\neq n}
\frac{
(f_m-f_n)V_{mn}^{(\alpha)}V_{nm}^{(\beta)}
}{
(E_m-E_n)(\hbar\omega+i\eta-E_n+E_m)
},
\label{eq:sigma_inter}
\end{align}
\end{widetext}
while the intraband term is
\begin{equation}
\sigma_{\alpha\beta}^{\mathrm{intra}}(\omega)
=
\frac{e^2\hbar}{i(2\pi)^2}
\int_{\mathrm{BZ}}d^2k
\sum_n
\left(
-\frac{\partial f}{\partial E_n}
\right)
\frac{
V_{nn}^{(\alpha)}V_{nn}^{(\beta)}
}{
\hbar\omega+i\eta
}.
\label{eq:sigma_intra}
\end{equation}
Here $f_n$ is the Fermi--Dirac distribution. In the non-Hermitian calculation we take the occupation as a function of the real part of the quasiparticle energy, $f_n=f[\mathrm{Re}(E_n)]$, which corresponds to a linear-response treatment in which the reservoirs fix the electronic occupation while the imaginary self-energy modifies lifetimes and optical matrix elements. The parameter $\eta$ is a phenomenological broadening that regularizes the response and represents residual scattering not explicitly included in $H_{\xi,\mathrm{NH}}$. The spin and valley degeneracies are included through
\begin{equation}
\sigma_{\alpha\beta}^{\mathrm{tot}}(\omega)
=
g_s\sum_{\xi=\pm1}
\sigma_{\alpha\beta}^{(\xi)}(\omega).
\end{equation}

This formulation differs from the Hermitian Kubo expression in two essential ways. First, the matrix elements are evaluated between left and right eigenstates rather than ordinary Hermitian eigenvectors. Second, the complex spectrum modifies the resonance denominators and therefore shifts both the position and linewidth of the collective modes. 

\section{Non-Hermitian Plasmons in Twisted Bilayer Graphene}
\begin{figure*}
		\centering
		\includegraphics[width=\linewidth]{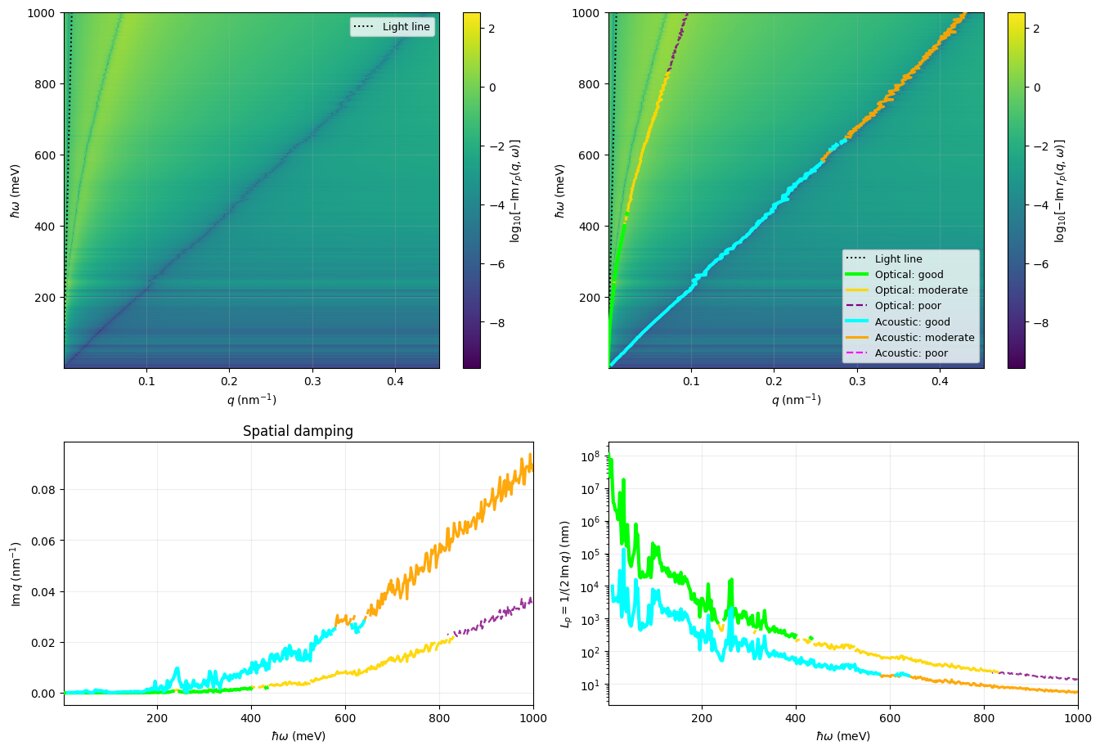}
		\caption{Upper panel: Plasmon dispersion and loss map of NH-TBG for gain $g=100~\mathrm{meV}$. Two collective modes are observed in the plasmonic regime, namely an optical branch and an acoustic branch with approximately linear dispersion. The acoustic branch is highly confined. The colormap represents the poles of $\log_{10}[-\text{Im}(r_p(q,\omega))]$. Lower panel: Spatial damping, left, and propagation length on a logarithmic scale, right. At low excitation energies, the imaginary part of the complex wave vector is strongly reduced in the ideal linear gain--loss model, producing very large formal propagation lengths. These values should be interpreted as upper bounds before including extrinsic disorder, substrate absorption, radiative leakage, and gain saturation.}
		\label{fig:modecomp}
	\end{figure*}
	
	\begin{figure*}
		\centering
		\includegraphics[width=\linewidth]{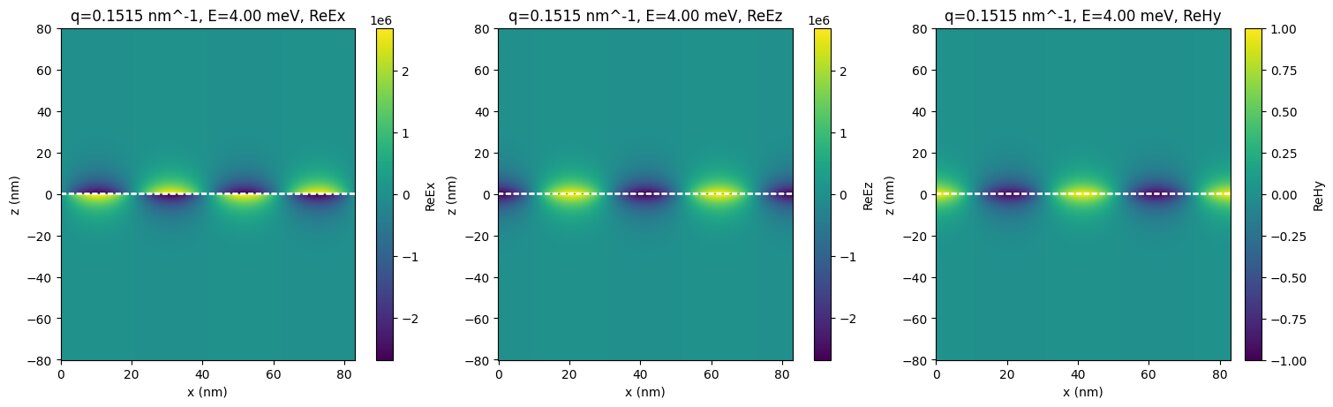}
		\caption{Representative transverse-magnetic field profiles of NH-TBG in a $\PT$-symmetric configuration with balanced gain and loss, $g_{\mathrm{gain}}=g_{\mathrm{loss}}=50~\mathrm{meV}$. The panels show the real parts of the dominant field components for the mode at the indicated $(q,\hbar\omega)$. The plotted amplitudes use the normalization of the field-profile calculation and should be interpreted as spatial mode profiles rather than an absolute passive--active amplitude calibration. The localization near the TBG plane follows from the large real plasmon wave vector and the reduced effective damping near the plasmon pole.}
		\label{fig:fieldcomp}
	\end{figure*}
	
	\begin{figure*}
		\centering
		\includegraphics[width=\linewidth]{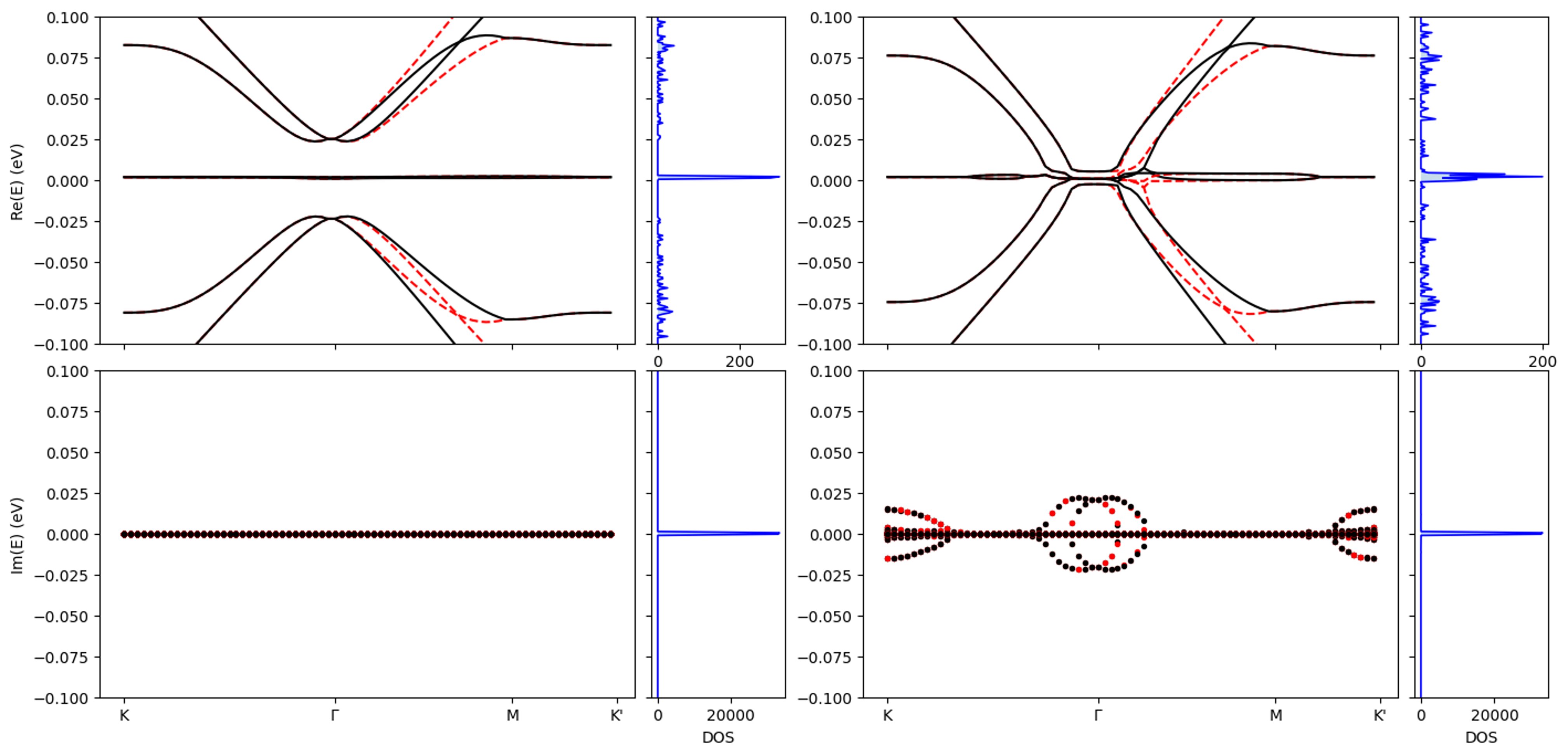}
		\caption{Electronic structure and density of states used to evaluate the conductivity kernel. Left: Passive calculation without the $\PT$-symmetric gain--loss term. Right: Calculation in the $\PT$-symmetric configuration with $\mathrm{gain}=\mathrm{loss}=50~\mathrm{meV}$. The upper panels show $\mathrm{Re}\,E$ along the moir\'e high-symmetry path, while the lower panels show $\mathrm{Im}\,E$. With black (red), we depict the valley $\xi = +1\hspace{2pt}(-1)$ of the electronic structure calculation. The density-of-states panels are obtained by binning the real and imaginary parts of the complex eigenvalues after diagonalizing the continuum Hamiltonian on the chosen moir\'e momentum grid.}
		\label{fig:nhbsdoscomp}
	\end{figure*}
	
\begin{table*}
\caption{Numerical simulations above the critical temperature of TBG for various non-Hermitian parameters.}
\label{tab:nhnumres}
\begin{ruledtabular}
\begin{tabular}{c|cccccccc}
$T(\mathrm{K})$
& $gain/loss\hspace{2pt}[eV]$
& $\dfrac{\Im}{\Re}(q)^{opt}_{max}$
& $\dfrac{\Im}{\Re}(q)^{ac}_{max}$
& $\eta_{min}^{opt}$
& $\eta_{max}^{opt}$
& $\eta_{min}^{ac}$
& $\eta_{max}^{ac}$ 
& $L_p=\dfrac{1}{2|\Im q|}\,[nm]$\\
&&&&&&&&\\
\hline
10 & 0.1 / 0 & 0.40 & 0.20 & 1.14 & 25.64 & 76.76 & 133.56 &  $10^7-10^8$\\
& 0.2 / 0 & 0.41 & 0.23 & 1.01 & 17.27 & 50.37 & 61.32 &  $10^6-10^7$\\
& 0.3 / 0 & 0.56 & 0.33 & 1.00 & 12.30 & 36.15 & 53.19 &  $10^8-10^9$\\
& 0.4 / 0 & 0.38 & 0.34 & 1.00 & 9.09 & 26.11 & 49.78 &  $10^9$\\
& 0.5 / 0 & 0.0054 & 0.22 & 1.00 & 1.27 & 9.89 & 11.09 &  $10^9-10^{10}$\\
\hline
50 & 0.1 / 0 & 1.16 & 0.58 & 1.10 & 24.64 & 76.54 & 110.90 &  $10^7-10^8$\\
& 0.2 / 0 & 1.35 & 0.92 & 1.0018 & 40.94 & 60.48 & 148.74 &  $10^8$\\
& 0.3 / 0 & 0.17 & 0.23 & 1.00 & 5.05 & 17.34 & 31.91 &  $10^8-10^{10}$\\
& 0.4 / 0 & 0.17 & 4.38 & 0.96 & 2.06 & 0.22 & 67.58 &  $10^8-10^9$\\
& 0.5 / 0 & 0.0026 & 0.30 & 1.00 & 1.02 & 4.48 & 6.91 &  $10^9-10^{10}$\\
\hline
100 & 0.1 / 0 & 6.62 & 16.48 & 1.61 & 19.52 & 1.90 & 305.16 &  $10^6$\\
& 0.2 / 0 & 1.31 & 0.90 & 1.0018 & 38.69 & 59.41 & 142.53 &  $10^7-10^8$\\
& 0.3 / 0 & 0.41 & 0.35 & 0.95 & 11.57 & 21.17 & 48.21 &  $10^8-10^{9}$\\
& 0.4 / 0 & 0.033 & 0.22 & 1.00 & 2.47 & 16.67 & 19.35 &  $10^9-10^{10}$\\
& 0.5 / 0 & 0.001 & 0.23 & 1.00 & 1.01 & 4.81 & 4.92 &  $10^{10}$\\
\hline
300 & 0.1 / 0 & 1.82 & 0.77 & 1.28 & 28.13 & 75.86 & 153.72 &  $10^5-10^6$\\
& 0.2 / 0 & 0.51 & 0.28 & 1.02 & 19.77 & 56.74 & 69.54 &  $10^7$\\
& 0.3 / 0 & 0.61 & 0.36 & 1.00 & 12.88 & 37.21 & 52.72 &  $10^7-10^{8}$\\
& 0.4 / 0 & 0.06 & 0.22 & 1.00 & 4.17 & 21.67 & 26.465 &  $10^9$\\
& 0.5 / 0 & 0.001 & 0.23 & 1.00 & 1.07 & 7.50 & 7.83 &  $10^9-10^{10}$\\
\hline\hline
10 & 0.05 / 0.05 & 5.68 & 3.02 & 1.17 & 79.64 & 58.80 & 313.25 &  $10^5-10^6$\\
50 & 0.05 / 0.05 & 3.34 & 1.00 & 1.79 & 170.78 & 21.31 & 827.21 &  $10^4-10^5$\\
100 & 0.05 / 0.05 & 6.15 & 1.20 & 0.74 & 49.89 & 10.58 & 820.07 &  $10^4-10^5$\\
300 & 0.05 / 0.05 & 5.30 & 1.28 & 0.09 & 50.55 & 7.80 & 219.61 &  $10^3-10^4$\\

\end{tabular}
\end{ruledtabular}
\end{table*}

The numerical workflow is as follows. For each choice of twist angle, tunneling amplitudes, temperature, chemical potential, broadening, dielectric environment, and gain--loss parameters, we diagonalize Eq.~\eqref{eq:HNH} on the moir\'e momentum grid and construct the right and left eigenvectors. The band structure is obtained by following the high-symmetry path $K-\Gamma-M-K'$ of the moir\'e Brillouin zone and plotting separately the real and imaginary parts of the complex eigenvalues (see below in Fig.~\ref{fig:nhbsdoscomp}). The density-of-states panels shown with the band structure are obtained from the same finite eigenspectrum by histogramming either the real or imaginary part of the complex eigenvalues,
\begin{equation}
\rho_X(x_j)=
\frac{1}{N_k\Delta x}
\sum_{\mathbf{k},n}
\chi_j\!\left[X_n(\mathbf{k})\right],
\qquad
X_n\in\{\mathrm{Re}\,E_n,\mathrm{Im}\,E_n\},
\label{eq:dos}
\end{equation}
where $\Delta x$ is the bin width and $\chi_j(y)=1$ when $y$ lies inside the $j$th histogram bin and zero otherwise. This is the finite-bin spectral distribution used in the plotting workflow, rather than a resolvent-broadened Green-function trace. The imaginary-eigenvalue histogram is included only to display the non-Hermitian spectral broadening or amplification associated with the $\PT$-symmetric environment; it is not interpreted as an additional real-energy band.

The numerical parameters used for this study are $\theta = 1.05^{\circ}$, $v_F = 10^6\hspace{2pt}\mathrm{m/s}$, tunneling amplitudes $w_0 = 79.7\hspace{2pt}\mathrm{meV}$ and $w_1 = 97.5\hspace{2pt}\mathrm{meV}$, $\eta = 2\hspace{2pt}\mathrm{meV}$, and $T\in [10,300]\hspace{2pt}\mathrm{K}$. The momentum-grid size used in the simulations for the optical conductivity was a $50\times50$ grid.

The plasmon modes are obtained from the poles of the electromagnetic response using the conductivity obtained from the non-Hermitian Kubo calculation. In the numerical implementation we solve the transverse-magnetic boundary-condition equations for the two-layer geometry. Defining the dimensionless variables
\begin{gather}
B=\frac{\hbar c q}{E_F},
\qquad
u=\frac{\hbar\omega}{E_F},
\qquad
\delta_d=\frac{E_F d}{\hbar c},\\
k_j=\left(B^2-\varepsilon_j u^2\right)^{1/2},\nonumber
\end{gather}
and the normalized conductivity $\tilde{\sigma}(\omega)=\sigma(\omega)/(e^2/h)$, the optical and acoustic branches are found from
\begin{align}
D_{\mathrm{opt}}(B,u)
=
1+
\frac{\varepsilon_2}{\varepsilon_1}
\frac{k_1}{k_2}
\tanh\!\left(\frac{\delta_d k_2}{2}\right)
+i\frac{2\alpha}{\varepsilon_1}
\tilde{\sigma}(\omega)
\frac{k_1}{u}
=0,
\label{eq:dielectric}\\
D_{\mathrm{ac}}(B,u)
=
1+
\frac{\varepsilon_2}{\varepsilon_1}
\frac{k_1}{k_2}
\coth\!\left(\frac{\delta_d k_2}{2}\right)
+i\frac{2\alpha}{\varepsilon_1}
\tilde{\sigma}(\omega)
\frac{k_1}{u}
=0.
\end{align}
Here $\alpha$ is the fine-structure constant, $E_F = 0.3\hspace{2pt}\mathrm{eV}$ is the energy scale used to nondimensionalize the wave vector, $d = 3.35\hspace{2pt}\text{\AA}$ is the interlayer separation, and $\varepsilon_1 = 1$, $\varepsilon_2 = 3.9$ are the dielectric constants on the two sides of the bilayer. The first equation corresponds to the optical plasmon branch, in which the layer charge densities oscillate predominantly in phase, while the second describes the acoustic branch, in which the charge oscillations are predominantly out of phase. The acoustic mode is especially sensitive to interlayer Coulomb coupling and therefore provides a natural probe of the moir\'e-modified low-energy electronic structure.

In the present work, we solve the plasmon condition at real excitation energy $\hbar\omega$ and obtain a complex wave vector
\begin{equation}
q(\omega)=q'(\omega)+iq''(\omega).
\end{equation}
The real part $q'$ determines the plasmon wavelength, while the imaginary part $q''$ determines spatial attenuation or amplification. In the numerical postprocessing the attenuation length is computed from the magnitude of the imaginary wave vector,
\begin{equation}
L_p(q)=\frac{1}{2|q''|},
\label{eq:prop_length}
\end{equation}
which reduces to the usual expression $1/(2q'')$ for a decaying mode written with $q''>0$. If the ideal linear model gives the opposite sign convention, the solution should be interpreted as an amplifying or threshold-like mode rather than as an indefinitely stable propagating plasmon. The confinement factor is
\begin{equation}
\eta(q)
=
\frac{\hbar c q'}{\sqrt{\varepsilon_{\mathrm{eff}}}\,\omega}
=
\frac{\lambda_0}{\lambda_p}.
\label{eq:confinement}
\end{equation}
Large values of $\eta(q)$ indicate strong subwavelength confinement. Compared to the Hermitian case ($\lambda_t=\lambda_b=0$), the non-Hermitian configuration can increase the formal propagation length while preserving strong confinement. The effective gain term may be approached experimentally through optical pumping or coupling to an active environment, but the present model does not include the nonlinear saturation, heating, and reservoir dynamics that would determine the ultimate device-level gain limit.

\section{Discussion of Results}

Figure~\ref{fig:modecomp} shows the plasmon dispersion and loss map of non-Hermitian TBG for gain strength $g=100~\mathrm{meV}$. Two collective branches are visible in the plasmonic regime: an optical branch and an acoustic branch. The optical branch displays the expected higher-energy character, while the acoustic branch exhibits an approximately linear dispersion and strong wave-vector confinement. The existence of optical and acoustic branches follows from the bilayer Coulomb problem, but their detailed dispersion and damping are controlled by the moir\'e conductivity entering Eq.~\eqref{eq:dielectric}. This is the point at which the twisted structure differs from a generic bilayer: the reconstructed bands and enhanced low-energy density of states modify the current matrix elements and the available intraband and interband transitions.

The mechanism responsible for the plasmon enhancement can be traced to the denominator of the plasmon condition. In the passive system, the imaginary part of the conductivity and residual scattering produce spatial attenuation. In the $\PT$-symmetric non-Hermitian configuration, the complex eigenvalues and biorthogonal velocity matrix elements modify both the real and imaginary parts of $\sigma(\omega)$. When the gain--loss contribution partly compensates the passive damping, the zero of $D_{\mathrm{opt}}$ or $D_{\mathrm{ac}}$ moves closer to the real-$q$ axis and the propagation length increases. Near $\PT$-symmetry-breaking thresholds or exceptional-point-like spectral degeneracies, this effect can be amplified because the left and right eigenvectors become strongly nonorthogonal and the optical matrix elements are redistributed. The present calculations should therefore be interpreted as evidence for $\PT$-induced spectral reshaping of the plasmon pole. A strict identification of an exact exceptional point would require an additional scan showing simultaneous coalescence of both eigenvalues and eigenvectors; we do not rely on such a claim here.

The corresponding electronic structure and density of states are shown in Fig.~\ref{fig:nhbsdoscomp}. The comparison between the passive and $\PT$-symmetric cases demonstrates that non-Hermiticity modifies the low-energy spectral weight of the moir\'e bands. The lower panels, which show $\mathrm{Im}\,E$, are important because they indicate where the $\PT$-symmetric environment produces finite lifetime or amplification channels in the electronic spectrum. Since the plasmonic response is controlled by the conductivity kernel, this redistribution of spectral weight directly affects both the plasmon dispersion and the damping rate. Thus, the observed enhancement is not simply an external electromagnetic gain effect, but results from the interplay between moir\'e band reconstruction and non-Hermitian spectral modification.

Figure~\ref{fig:fieldcomp} presents transverse-magnetic field profiles in the $\PT$-symmetric configuration with balanced gain and loss, $g_{\mathrm{gain}}=g_{\mathrm{loss}}=50~\mathrm{meV}$. The plotted fields are normalized mode profiles. The dominant physical reason for the localization is the large real plasmon momentum $q'$, which produces evanescent decay away from the TBG plane. The gain--loss term then reduces the effective attenuation near the plasmon pole, so the field can remain strongly confined while suffering less spatial decay. Therefore the field figure should be used to support localization and modal structure; a quantitative passive--active amplitude enhancement would require a separate comparison using a common field normalization.

The numerical trends summarized in Table~\ref{tab:nhnumres} further support this interpretation. Increasing the gain/loss parameter generally changes both the confinement factor and the attenuation ratio $\mathrm{Im}\,q/\mathrm{Re}\,q$. In several parameter regimes, the acoustic mode combines strong confinement with reduced damping, making it the most promising branch for non-Hermitian plasmonic functionality. However, the response is not monotonic in either temperature or gain/loss strength. This nonmonotonicity is expected in TBG because small changes in spectral weight near the reconstructed moir\'e bands can strongly affect the Kubo conductivity.

The very large propagation lengths reported in Fig.~\ref{fig:modecomp} and Table~\ref{tab:nhnumres} should be read as ideal upper bounds of the linear model. Mathematically, $L_p=1/(2|q''|)$ becomes very large whenever gain compensation drives $|q''|$ close to zero. In an experimental device, this divergence would be cut off by disorder, substrate absorption, radiative leakage, carrier relaxation, inhomogeneous twist-angle broadening, finite sample size, heating, and nonlinear gain saturation. If the gain overcompensates loss and produces a spatially growing solution under the chosen sign convention, the mode is amplifying rather than a stable low-loss plasmon. Therefore the relevant experimental target is not infinite or macroscopic propagation, but a regime in which residual damping is reduced while the mode remains below the instability threshold.

This stability requirement is central to the $\PT$-symmetric interpretation. In the exact $\PT$-symmetric regime, the spectrum can remain effectively real over part of parameter space, and gain and loss are balanced at the level of the linearized mode. At the $\PT$-broken threshold, pairs of modes acquire opposite imaginary parts; beyond this point, one mode may grow unless nonlinear saturation or external loss clamps the amplification. The stable operating window is therefore bounded by passive damping on one side and net-gain instability on the other. Disorder, substrate loss, and imperfect gain--loss balance will shift this window and smear any sharp exceptional-point signatures.

Importantly, all calculations are performed above the superconducting critical temperature of TBG. Therefore, the predicted enhancement is not attributed to superconducting collective modes or condensate stiffness. Instead, it arises from the normal-state moir\'e electronic structure combined with gain--loss-induced modification of the optical conductivity. This distinction is important because it suggests that non-Hermitian plasmonic control may be achievable without requiring superconducting order.

\section{Conclusion}

We have investigated plasmonic excitations in twisted bilayer graphene using a non-Hermitian extension of the Bistritzer--MacDonald model combined with a biorthogonal Kubo formalism. This approach allows the optical conductivity and plasmonic response to be computed in the presence of effective gain, loss, and complex electronic spectra.

Our results show that non-Hermitian TBG supports both optical and acoustic plasmon branches. The acoustic branch is strongly confined and displays a nearly linear dispersion, making it particularly sensitive to interlayer Coulomb coupling and moir\'e-band reconstruction. In $\PT$-symmetric gain--loss configurations, the effective plasmon damping can be substantially reduced within the ideal linear model, leading to enhanced formal propagation lengths and localized transverse-magnetic field profiles.

The enhancement reported here should be understood as a gain-assisted non-Hermitian effect in the normal metallic state of TBG. It does not rely on superconductivity, but instead originates from the combined influence of moir\'e electronic structure, biorthogonal optical transitions, and gain--loss-modified plasmon poles. The results are consistent with $\PT$-induced spectral reshaping and possible proximity to exceptional-point-like regimes, but an exact exceptional point would require an additional eigenvector-coalescence analysis.

These findings identify twisted bilayer graphene as a promising platform for active non-Hermitian plasmonics. More broadly, they establish a connection between moir\'e quantum materials, graphene plasmonics, and $\PT$-symmetric wave physics. Future work should include disorder, substrate loss, nonlinear gain saturation, finite-size effects, and a full stability analysis of the complex plasmon spectrum.

\section*{Acknowledgments}
This research was co-funded by the Stavros Niarchos Foundation (SNF) and the Hellenic Foundation for Research and Innovation (H.F.R.I.) under the 5th Call of Science and Society Action---Always Strive for Excellence---Theodore Papazoglou (Project Number: 011496).

\section*{Data availability}
The numerical data and code that support the findings of this study are available from the corresponding author upon reasonable request. No experimental datasets were generated in this work.

\section*{Conflict of Interest}
The authors declare no competing financial interests or personal relationships that could have appeared to influence the work reported in this paper.

\bibliographystyle{apsrev4-2}
\bibliography{nh_plasmonics_tbg_refs_revised}
\end{document}